\documentclass[journal]{IEEEtran}
\usepackage{amssymb}
\usepackage{tabularx}
\usepackage{amsmath}
\usepackage{graphicx}%
\usepackage{cite}

\begin{document}

\title{Maximum Q-factor of planar inductors}

\author{{Mohamed~Ismail~Abdelrahman,~\IEEEmembership{Graduate Student~Member,~IEEE,}
        Matteo~Ciabattoni,~\IEEEmembership{Graduate Student~Member,~IEEE,}
        and~Francesco~Monticone,~\IEEEmembership{Senior Member,~IEEE}
}
\thanks{The Authors are with the School of Electrical and Computer Engineering, Cornell University, Ithaca, NY, 14853 USA. (email: francesco.monticone@cornell.edu)}}

\maketitle

\begin{abstract}
On-chip inductor design plays a critical role in the advancement of radio-frequency integrated circuits (RFICs). Inductors typically occupy a substantial portion of the chip area as their performance metrics, namely, inductance density and Quality factor ($Q$-factor), are fundamentally tied to the available footprint, thereby limiting miniaturization. To better understand and quantify these limitations, we employ rigorous electromagnetic analysis together with convex optimization techniques to derive a fundamental bound on the maximum achievable $Q$-factor of electrically-small planar inductors as a function of the available design area. 
The analysis yields analytical expressions for the bound and, via modal analysis techniques, identifies and interprets operational regimes and scaling trends with respect to design area and material conductivity. The analysis accounts for both ohmic and radiation losses, with the latter becoming significant as the inductor size increases. A broad set of state-of-the-art inductor designs from the literature is evaluated against the established $Q$-factor upper bound, identifying designs that approach the theoretical limit as well as those with potential for further improvement. The study is extended to include the effect of kinetic inductance, which offers a promising avenue toward next-generation inductors with higher inductance densities and $Q$-factors. By establishing this benchmark, this work aims to guide and inspire the design of more efficient and compact planar inductors for high-performance RF systems.
\end{abstract}


\IEEEpeerreviewmaketitle

\section{Introduction}
\IEEEPARstart{A}{dvanced} electromagnetic and nanophotonic devices increasingly rely on inverse design techniques, which are highly effective in discovering unconventional designs that can meet challenging performance objectives, such as maximizing absorption, achieving specific scattering patterns, or enabling efficient light-matter interactions \cite{molesky2018inverse,pestourie2018inverse,li2022empowering,jiang2021metasurface}. A major limitation of standard inverse design techniques is that they do not account for the system’s fundamental performance limits. This can often result in inefficient use of computational resources, particularly when the algorithm attempts to meet design objectives that, unknowingly, exceed the physical limits of the system. In this context, \emph{a priori} knowledge of fundamental performance bounds is highly valuable for guiding computational methods by setting achievable limits for a given design objective, thereby narrowing the search space and significantly reducing the computation time \cite{kuang2020maximal,gustafsson2020upper,kuang2020computational,chao2022physical, strekha2024limitations,  amaolo2024maximum,chao2025bounds, li2025approaching,gertler2025many}.

The development of analytical bounds in electromagnetic designs can be traced back to the work of Wheeler and Chu on electrically-small antennas \cite{wheeler1947fundamental,chu1948physical}. These pioneering studies established a fundamental relationship between the physical size of a small antenna and its maximum achievable bandwidth (or minimum quality factor, $Q_{min}$). Minimizing the $Q$-factor, defined as the ratio of the energy stored in a system to the energy dissipated per cycle, in addition to reducing ohmic losses, are crucial to realize broadband and energy-efficient radiating systems. In Chu's original analysis, only the fields radiated outside the smallest sphere of radius $a$ enclosing the sources were considered, leading to an analytical expression for the bound: $Q_{min}=(ka)^{-3}$, where $k$ is the wavenumber in free space. This spherical approximation, however, leads to a \textit{loose} bound (overly optimistic) when applied to antennas that do not fill a spherical volume, such as wire-based, cylindrical, or planar designs commonly used in wireless communication devices \cite{gustafsson2009illustrations}. As a result, the Chu bound is unattainable for any conceivable implementation of these configurations \cite{mohammadpour2011physical}. To overcome this limitation, researchers have extended the analysis to account for non-spherical optimizable geometries, including planar antenna structures, leading to more practical (i.e., tighter) bounds \cite{mohammadpour2011physical,foltz1999limits,thal2006new,gustafsson2009illustrations,hansen2011minimum,yaghjian2010lower}. In this context, a \textit{tight} bound indicates that there exists at least one physically realizable design within the optimizable area/volume that approaches the bound value.

More broadly, in recent years, research efforts have increasingly expanded beyond canonical geometries, towards the development of a systematic framework for establishing ultimate performance bounds that can be applied to more practical configurations, including electrically-large structures. Advanced techniques such as convex optimization relaxations and eigenvalue-based methods have been instrumental in deriving a wide array of performance bounds across a broad range of electromagnetic applications, including the determination of $Q_{min}$ for arbitrary-shaped antennas, maximum antenna gain $Q$-factor quotient, maximum spectral efficiency for MIMO antennas, bounds on plane-wave scattering and absorption in thin-films, optimal metasurface reflection across arbitrary angles and polarizations, and many others  \cite{gustafsson2016antenna,gustafsson2012optimal,chalas2016computation,capek2019optimal,ehrenborg2020physical,kuang2020maximal,gustafsson2020upper,shim2021fundamental,li2024approaching,abdelrahman2023thin,nel2024radiation,chao2022physical,gertler2025many}.

Although $Q_{min}$ bounds have been the subject of many studies over the past decades, especially in the context of optimal antenna performance, a fundamental problem that has been largely overlooked in the literature is the complementary problem of determining the \emph{upper} bound on the $Q$-factor, particularly for magnetic systems ($Q^m_{max}$), namely, systems that store most of their energy in the magnetic field. The $Q$-factor of magnetic systems is known to be fundamentally bounded, even at small sizes \cite{best2004lower}, unlike electric systems for which the $Q$-factor can,  in principle, be unbounded. Another fundamental property of magnetic systems is that their inductance and $Q$-factor values are intrinsically tied to the physical size, as larger areas allow for a greater magnetic flux to be generated and stored. Consequently, the design of high-$Q$ inductors with a small footprint is a notoriously challenging task.

Maximizing the $Q$-factor of inductors is crucial for a wide range of applications that require a selective bandwidth (resonant) response, or strongly confined fields with minimal energy dissipated through heat loss or radiation. For example, the efficiency of inductively-coupled wireless power transfer systems is directly correlated with the $Q$-factor of the inductors,  enabling strong and long-range coupling between transmitter and receiver circuits \cite{pham2021optimal}. In RFICs, high-Q inductors are the backbone of low-loss matching networks, narrowband filters, and low phase-noise oscillators \cite{borwick2003high}.
The sharp spectral selectivity associated with high-Q resonators is also essential for high-resolution sensing applications, such as planar RFID sensors that rely on resonant structures to encode data or detect environmental variations
\cite{menon2022development,lasantha2025trade}. Furthermore, achieving a high $Q$-factor is critical for maintaining long coherence times in superconducting qubits, which exploit the nonlinear inductive properties of Josephson junctions \cite{paik2011observation}, as well as in the ultra-sensitive kinetic inductance detectors used in radio astronomy \cite{hammer2008ultra}. As these technologies move toward further miniaturization and integration, the challenge of realizing inductors with large inductance and a high $Q$-factor in limited physical spaces becomes increasingly important and difficult \cite{kang2018chip,park2022effective,hacohen2023flexible}. For these reasons, a fundamental upper bound $Q^m_{max}$ would serve as an indispensable tool for the technologically important task of designing high-performance planar inductors for a wide range of classical and quantum applications.

\begin{figure}[ht]
 \centering
\includegraphics[width=0.5\textwidth]{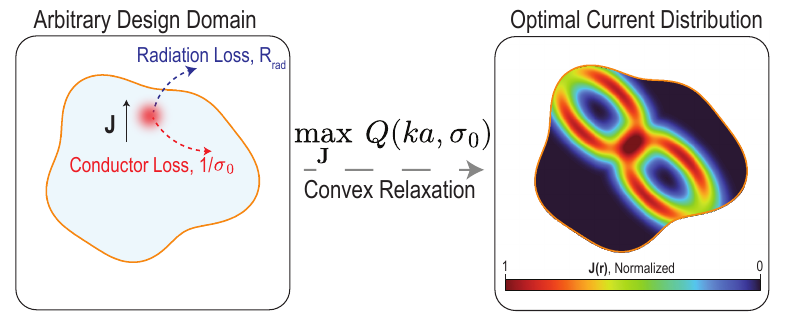}
\caption{\label{fig:fig0} Illustration of the problem of establishing upper bounds on the $Q$-factor of planar inductors within an arbitrary design domain, in the presence of both radiation and ohmic losses. While the original problem of finding the current distribution that maximizes $Q$ is non-convex, it can be transformed into a related convex problem whose optimal solution provides a bound to the original optimization problem. 
}
\end{figure}

Although there has been considerable research on the design and optimization of inductors under specific geometric and size constraints \cite{islam2013design,park2022effective}, the problem of identifying a fundamental upper limit for their $Q$-factor remains unresolved. 
Establishing this upper bound will not only deepen our understanding of the intrinsic limitations of inductors but also provide a rigorous benchmark to guide efforts toward more efficient inductor design while reducing computational efforts. The only existing work relevant to this problem, Ref. \cite{gustafsson2019tradeoff}, showed that planar inductors confined to a square area are optimal (compared to rectangular areas) to achieve the highest $Q$-factor per unit area. This insight, while valuable, does not establish a definitive upper bound for the maximum $Q$-factor achievable within an arbitrary design area. In this paper, we address this gap by presenting analytical expressions for an upper bound on the $Q$-factor of planar inductors, guided by a rigorous convex optimization framework, as illustrated in Fig.~\ref{fig:fig0}. We believe that our approach and results help bridge the gap between fundamental theoretical limits and practical design considerations for this technologically important problem. 

\section{Upper bounds on the $Q$-factor of planar inductors}\label{sec2}
Planar structures can be efficiently analyzed using the Method of Moments (MoM). This numerical technique discretizes the unknown induced surface current distribution on the structure into a finite set of basis functions. In this work, we use the Rao-Wilton-Glisson basis functions \cite{jenn2005radar} that represent the surface current as a superposition of elementary current dipoles, forming the vector $\bf{I}$. The MoM transforms the integral equations that govern the electromagnetic behavior of the structure into a system of linear equations given in matrix form, $\bf{V=Z\,I}$, where $\bf{V}$ represents the excitation, and the impedance matrix $\bf{Z}=\bf{R}+j\bf{X}$ encapsulates all the interactions between the induced elementary dipoles in the structure. $\bf{R}$ is the resistance matrix, $\bf{X}=\bf{X_m}-\bf{X_e}$ is the reactance matrix, and $\bf{X_m}$ and $\bf{X_e}$ are stored magnetic and electric energy matrices, respectively (see Appendix A). The surface current distribution can then be estimated numerically using matrix inversion techniques to find the coefficients of the basis functions. The impedance matrix $\bf{Z}$ is independent of the external excitation, which significantly improves computational efficiency, as the matrix only needs to be computed once for a given structure \cite{gibson2021method}.

The MoM framework is well-suited for deriving physical bounds in electromagnetism, as it discretizes the relevant integral equations and expresses key physical quantities, such as dissipated power, stored energy, and the $Q$-factor, as linear or quadratic functions of the current vector $\bf{I}$, written in the form of vector-matrix products (see Appendix A). These mathematical forms are amenable to established optimization techniques aimed at identifying a unique global optimum, namely, a fundamental performance bound \cite{gustafsson2020upper,liska2021fundamental}. In particular, the maximization problem for the $Q$-factor, $Q^m$, of an inductive structure [Eq. (\ref{eq:opt}) in Appendix B] represents a quadratically constrained quadratic program (QCQP) that is, however, non-convex. Consequently, this problem may admit multiple locally optimal solutions, making it challenging to identify the true global bound. Such challenges are common in many design optimization problems arising in electromagnetism and photonic design \cite{gertler2025many}. The standard approach to address this challenge is to transform such a non-convex problem into a related convex problem whose optimal solution could serve as a bound to the original optimization problem. In this paper, we adopt a computationally efficient method to tackle QCQPs known as semidefinite relaxation (SDR), which is polynomial in the problem size \cite{luo2010semidefinite,jonsson2017methods}. This method transforms the original nonconvex problem into a convex semidefinite program. For a broad class of problems, including the case of interest here, it can be shown that solving the semidefinite relaxation is equivalent to solving the original problem, in the sense that it yields the same globally optimal solution. In the following, we present the main results obtained with this approach and refer the reader to Appendix B for a detailed discussion.

\begin{figure*}
 \centering
\includegraphics[width=0.85\textwidth]{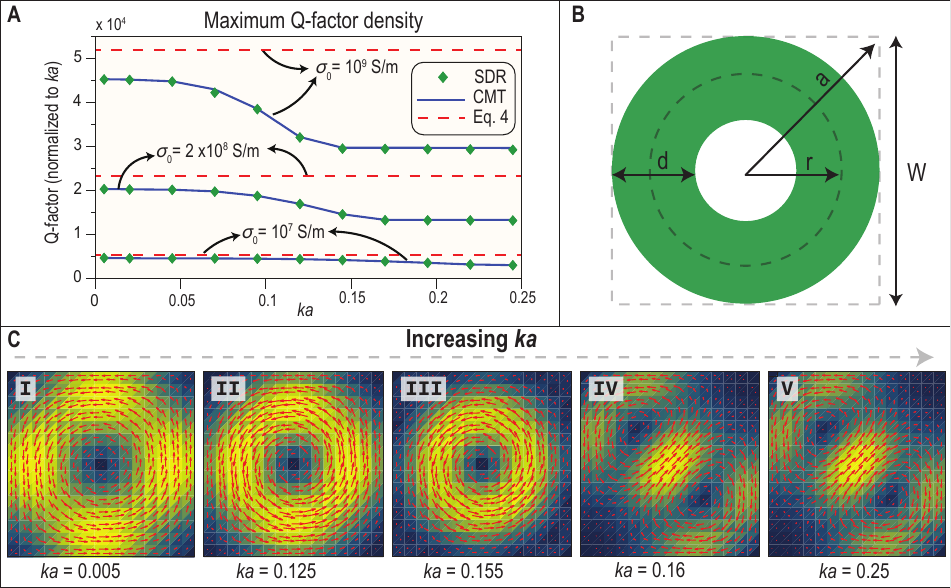}
\caption{\label{fig:fig1}  \textbf{(A)} Upper bound $Q^m_{\text{max}}$ (normalized to electrical size $ka$) for inductors confined to a square optimizable area of radius $a$, plotted as a function of electrical size and metal conductivity. The bound is computed using both convex optimization (SDR) and modal analysis (CMT) methods, as discussed in the Appendix B and C respectively, showing virtually perfect agreement. We have verified that the numerical solutions of $Q^m_{max}$ converge with increasing mesh resolution. The rigorous numerical bounds are compared with the approximate analytical bound given by Eq. (\ref{eq:Qm_max}) (green dashed curve). \textbf{(B)} Geometry used to derive the analytical formula for $Q^m_{\text{max}}$, motivated by the observed similarity between the optimal current distribution for small design areas and a single-turn circular loop with uniform current. The formula provides a closed-form, slightly loose upper bound for $Q^m_{\text{max}}$, in good agreement with results of optimization algorithms for small sizes, demonstrating its reliability as a benchmark for miniaturized inductor design where radiation losses are negligible. 
\textbf{(C)}  Optimal current distributions for different sizes of the optimizable design area, for $\sigma_0=2\times10^8$ S/m, illustrating the transition from a single wide-loop solution for small inductors to a two-turn solution for larger sizes. Configuration (iii), a single loop with reduced width, represents an intermediate stage where radiation losses begin to be significant. The onset of this transition strongly depends on the conductivity, as discussed in the main text.  In all sub-panels of \textbf{(c)}, brighter regions indicate a higher current density, while red arrows depict surface current magnitude and direction.
}
\end{figure*}

Figure \ref{fig:fig1}(a) shows the numerical solutions for $Q^m_{max}$, the rigorous upper bound on $Q$-factor for all possible planar inductor designs within an optimizable square area, plotted as a function of electrical size, $ka=2\pi a/\lambda$ ($a$ defined in Fig. \ref{fig:fig1}(b)), and metal conductivity $\sigma_0$. Here, to better highlight the non-trivial dependence between size and $Q$-factor, $Q^m_{max}$ is normalized by $ka$; hence, we plot the maximum $Q$-factor \emph{density}. The convex optimization results, obtained with the SDR method outlined above, are also compared with a modal analysis based on the characteristic modes theory (CMT) described in Appendix C section, showing excellent agreement. 

Three distinct regimes emerge depending on the size of the design area relative to the operating wavelength. For very small areas, $Q^m_{max}$ increases linearly with area size, corresponding to a constant $Q$-factor density that solely depends on the material conductivity.  As expected, higher conductivity (and thus lower losses) results in a higher $Q$-factor density bound. In this region, the optimal current distribution takes the form of a single loop of nearly uniform amplitude, as shown in  Fig. \ref{fig:fig1}(c).I-II. The relatively large width of the current loop ensures minimized ohmic resistance, while the small overall size of the current yields small radiation. Interestingly, the optimal current path avoids the square corners, reducing abrupt changes in the current direction, which is known to induce higher radiation losses \cite{balanis2015antenna}.

Analytical expressions for the inductance of circular loops \cite{balanis2015antenna} can provide valuable insight into these numerical results, due to the similarity between the optimal current distribution at small sizes and the current supported by a metallic loop (geometry shown in Fig. \ref{fig:fig1}(b)). The formula for the circular loop inductance in standard textbooks (e.g., Ref. \cite{balanis2015antenna}) can be used, after being adapted for 2D MoM analysis (see section 4.3 in Ref. \cite{makarov2002antenna}), resulting in
\begin{equation}
    L_{loop} = \mu_0 r \left[ \ln\left(\frac{8r}{d/2\pi}\right) - 2 \right],
\end{equation}
where $d$ is the loop width and $r$ is the loop radius, defined at the center of the loop width, as illustrated in Fig. \ref{fig:fig1}(b). The loop resistance is given by
\begin{equation}
    R_{loop} = \dfrac{2\pi r}{d} \, R_s, 
\end{equation}
where $R_s$ is the conductor surface resistance, which is inversely proportional to its conductivity (see Appendix A).

Radiation resistance can be neglected for small current loops, as it scales with $(ka)^4$ \cite{vandenbosch2010reactive}. Consequently, for the non-dispersive off-resonance regime characterized by small $ka$, the $Q$-factor of the circular loop  in Fig. \ref{fig:fig1}(b) can be expressed as a function of the inductor width $d$ as
\begin{equation} \label{eq:Qm_d}
 Q^m(d)= \dfrac{\omega   L_{loop}}{ R_{loop}} = \omega\dfrac{\mu_0}{2\pi R_s} d \left[ \ln\left(\frac{8\pi\, (W-d)}{d}\right) - 2 \right].
\end{equation}
where we have used $r=(W-d)/2 $, with $W=a\sqrt{2}$ denoting the side length of the square area. This analytical formula offers  valuable physical insights into the trade-off that determine the maximum possible value of $Q$-factor. Specifically, an increase in inductor width $d$ reduces the ohmic resistance, but also decreases the inductance due to a reduction in effective magnetic flux linkage. However, the inductance decreases logarithmically, and as a result, an optimal loop width exists. This optimum can be estimated by examining Eq. (\ref{eq:Qm_d}), which shows that the maximum $Q^m$ occurs at $d^*\approx0.4W$. The obtained optimal width, $d^*$, indeed closely matches the width of the numerically calculated optimal current distributions in Fig. \ref{fig:fig1}(c).I-II.

Substituting $d^*$ into Eq. (\ref{eq:Qm_d}), the upper bound on $Q^m$ can be estimated as
\begin{equation}\label{eq:Qm_max}
 Q^m_{max}\approx 0.15 \dfrac{ka}{R_s/\eta_0},
\end{equation}
where $\eta_0$ is the free space impedance. This formula provides an important and practically relevant result: a simple analytical expression for the upper bound on the $Q$-factor of small planar inductors, which depends only on the available design area, the operating frequency, and material losses, irrespective of the details of the inductor design. By plotting it in Fig. \ref{fig:fig1}(a), we see that the analytical bound (\ref{eq:Qm_max}) predicts slightly higher values compared to the more accurate numerical bound for small $ka$. The discrepancy can be attributed to the contribution of radiation losses neglected in Eq. (\ref{eq:Qm_d}), which becomes more significant at large sizes, as discussed in the next paragraph. As material losses increase towards practical values (for example, copper has $\sigma_0 = 5.8 \times 10^7$ S/m), ohmic losses become dominant and, as seen in Fig. \ref{fig:fig1}(a), the analytical formula more closely matches the numerical bound. The small error introduced by this simplification could be acceptable, as Eq. (\ref{eq:Qm_max}) remains a valid upper bound for all numerical results in Fig. \ref{fig:fig1}, and because radiation losses are negligible for most on-chip inductors. 

A different regime emerges for larger optimizable areas, as shown in Fig. \ref{fig:fig1}(c).IV-V, characterized by a different optimal current distribution with a two-loop configuration. This can be explained by the increased contribution of radiation losses at large sizes, which can no longer be neglected, as radiation losses of small loops grow much faster with size ($\propto a^4$) than ohmic losses ($\propto a$). Consequently, the optimal solution mitigates this effect by splitting the main loop into two smaller sub-loops carrying opposite current. This significantly reduces radiation losses by decreasing the radiating loop area (hence, the radiation resistance of each loop) and the overall magnetic dipole moment of the structure. While the two-loop configuration decreases radiation, it also increases ohmic losses, compared to a single wide loop, as a result of narrower and effectively longer current paths. Despite this trade-off, the two-loop configuration is the optimal solution for $Q^m_{max}$ for larger inductors, minimizing the combined effect of radiation and ohmic losses.
However, as shown in Fig. \ref{fig:fig1}(a), this trade-off leads to a maximum possible $Q$-factor density ($Q^m_{max}/ka$) that is lower than in the very-small-size regime, and that remains approximately constant as a function of $ka$.

Interestingly, Fig. \ref{fig:fig1} also shows that there exists a third, intermediate regime, where the numerical bound $Q^m_{max}/ka$ varies rapidly with $ka$ and the optimal current distribution remains a single-turn loop (Fig. \ref{fig:fig1}(c).III), albeit with a reduced width to decrease radiation loss. As decreasing the loop width becomes insufficient to counteract the rapidly increasing radiation losses, the optimal current distribution undergoes a transition and splits into the two-loop configuration described above. The transition to the larger-area regime is associated with a rapid variation in the numerical bound value. 
The size at which this transition occurs strongly depends on the conductivity value. Notably, the higher the conductor losses, the larger the size at which the impact of radiation losses becomes significant enough to cause the transition. For example, the reduced-width optimal current loop in Fig. \ref{fig:fig1}(c).III, which occurs just before the transition, is observed at $ka= 0.22, 0.15, 0.12$ for $\sigma_0= 10^7, 2\times10^8, 10^9$ S/m, respectively. These findings reveal a non-trivial relationship between the conductor properties and the optimal operating regime of inductors for achieving maximum $Q$-factor density.

\begin{figure}
\centering
\includegraphics[scale=0.65]{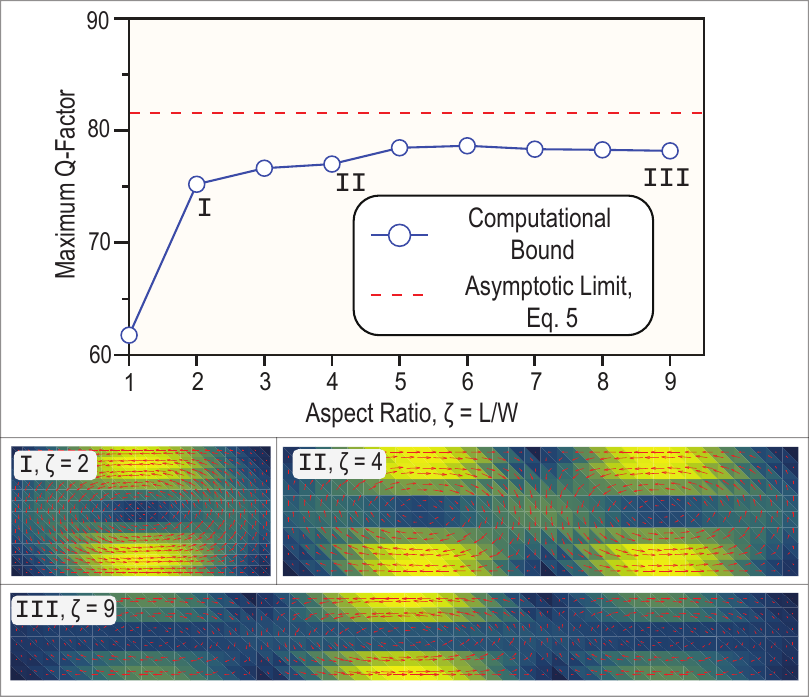}
\caption{\label{fig:fig2} Upper bound $Q^m_{\text{max}}$ (blue curve) numerically evaluated for rectangular design areas with a fixed width, $W=\lambda/50$, as a function of axial ratio. The results show that only modest growth is obtained as the length $L$ is increased. The corresponding optimal current distributions at different axial ratios $\zeta=L/W$ are shown in the insets (not to scale for clarity). 
The optimal current solution maintains a wide elliptical profile at moderate axial ratios, splitting into multiple smaller loops as $\zeta$ increases. Ultimately, the optimal current distribution converges to that found at moderate axial ratios, but repeated across the structure, explaining the observed plateau in $Q^m_{\text{max}}$. The saturation of $Q^m_{\text{max}}$ with increasing $\zeta$ implies that square or circular design areas are optimal for maximizing the $Q$-factor \emph{density} of inductors. In all cases, we assumed conductivity $\sigma_0=2 \times 10^8~\text{S/m}$ and a thickness much smaller than the skin depth. The approximate analytical formula for the bound, given by Eq. (\ref{eq:Q_max_rect}) for rectangular areas, is also plotted for comparison (red dashed line). }
\end{figure}

To generalize our analysis, the optimizable domain to achieve $Q^m_{max}$ is extended to include rectangular areas, as shown in Fig. \ref{fig:fig2}. Similar to the square case, for very small areas, the optimal current distribution corresponds to the widest elliptical loop that fits within the rectangular dimensions, effectively minimizing ohmic losses. Notably, starting from a square design area and increasing only one side (creating a rectangle with length-to-width ratio $\zeta=L/W>1$) can only lead to a moderate increase of $ Q^m_{max}$, as the inductance of a rectangular loop increases with $\zeta$ faster than the loop resistance \cite{jia2016finite}. As seen in Fig. \ref{fig:fig2}, the increase in $Q$-factor is not directly proportional to the increase in area; for instance, doubling the length of one side of a square inductor (leading to $\zeta=2$), yields an increase of the $Q$-factor by only about $25\%$. This implies that the highest achievable $Q$-factor \emph{density}, $Q^m_{max}/ka$ (with $2a=\sqrt{W^2+L^2}$), is obtained for square or circular design areas, as  reported in Ref. \cite{gustafsson2019tradeoff}.

Then, similar to the trend observed in Fig. \ref{fig:fig1}, as the area increases, driven by an increase in length-to-width ratio ($\zeta=L/W$) for a fixed width, the main elliptical current loop splits into two or more smaller loops (insets of Fig. \ref{fig:fig2}), to minimize the radiated power. 
The modest increase in $Q$-factor eventually saturates as $\zeta$ increases further, because the optimal current distribution converges to that found at moderate axial ratios, merely replicated across the structure. This behavior ultimately results in a significant reduction in $Q$-factor density. In this sense, if the goal is to optimize $Q$-factor density rather than absolute $Q$-factor, the preference for compact and symmetric shapes arises from their ability to effectively minimize the sum of both ohmic and radiative losses within a small footprint, ensuring efficient inductor design. 

This finding suggests that the saturated $Q^m_{max}$ for large aspect ratios is primarily governed by the smallest dimension of the rectangular design area. By simple fitting, the analytical bound formula in Eq. (\ref{eq:Qm_max}) can then be modified to predict this \textit{asymptotic} limit at large aspect ratios for rectangular design areas, as shown in Fig. \ref{fig:fig2}, resulting in 
\begin{equation}
\label{eq:Q_max_rect}
 Q^m_{max}\approx 0.125\, \dfrac{k\, \text{min}(W,L)}{R_s/\eta_0}.
\end{equation}

Finally, we note that, for planar and on-chip inductors, there are other factors beyond maximizing the $Q$-factor (or its density) that must be considered for optimal circuit performance. Notably, maximizing the inductance density, defined as inductance per unit area, is required for next-generation ICs \cite{kang2018chip}. A common approach for achieving high inductance density is to use multi-turn conductor designs, which leverage the mutual inductance contributions between neighboring turns. This results in a quadratic increase in inductance with the total number of turns \cite{mohan1999simple}. However, adding more turns within a fixed area means a reduction in the conductor width for each turn and an increase in the total conductor length, both leading to higher ohmic losses. This trade-off intuitively explains why the $Q$-factor is ultimately controlled by the overall available area for the inductor. This fact poses a significant challenge in optimizing the $Q$-factor of a multi-turn inductor. An optimal number of turns for achieving $Q^m_{\text{max}}$ has been identified for specific inductor layouts \cite{park2022effective}, while various innovative designs have been proposed to improve the $Q$-factor of planar inductors \cite{islam2013design,park2022effective,awuah2023novel}. 

In this context, while the optimal current distributions $I_{opt}$ shown in Fig. \ref{fig:fig1}(c) (optimal with respect to $Q^m$) 
might not correspond to ideal designs for practical inductors, the numerical and analytical bounds established here can definitely be helpful in assessing whether a given design is already very close to its maximum possible Q-factor, given its area and material, or if further improvements are possible. In Table \ref{tab:inductor_bounds}, we show this comparison for various existing planar inductor designs. In this comparison, the current cross-section is assumed to be limited in thickness due to skin depth by $2\delta$, to account for the effect at both conductor surfaces.Eq. (\ref{eq:Qm_max}) can then be rewritten as a $Q$-factor density bound that is fully independent of geometry,
\begin{equation}\label{eq:Qm_maxMax}
 Q^m_{max}/ka= 0.3\,\eta_0\, \sigma_0\, \delta.
\end{equation}

\begin{table}[h]  
    \caption{A comparison of the $Q$-factor upper bound for planar inductors, Eq. (\ref{eq:Qm_maxMax}), with the reported $Q$-factor of several designs from the literature across different frequency ranges.}
    \begin{tabularx}{\columnwidth}{|X|c|c|c|}
        \hline
        \textbf{Inductor Design} & \textbf{Operating frequency} & \textbf{$Q$-factor} & \textbf{$Q^m_\text{max}$} \\
        \hline
        \cite{awuah2023novel} Rectangular  coil & 100 KHz & 48.3 & 134 \\
        \cite{ben2023model} Spiral coil & 100 KHz & 46 & 57 \\
        \cite{park2022effective} Spiral coil & 2 MHz & 410 & 1302 \\
        & 6.78 MHz & 744 & 2397 \\
        \cite{kim2017optimal} Circular wire coil & 6.78 MHz & 148 & 240 \\
        \cite{kaziz2020tuning} PCB inductor & 42 MHz & 352 & 843 \\
        \cite{ma2023high} Flex-compatible inductor & 2.4 GHz & 65 & 463 \\
        \cite{pudari2024design} Fractal inductor & 7 GHz & 15 & 69 \\
        \cite{hashim2021analysis} Circular inductor & 10.6 GHz & 89 & 160 \\
        \hline
    \end{tabularx}
    
    \label{tab:inductor_bounds}
\end{table}
\section{Effect of Kinetic inductance}\label{sec3}
As discussed in the Introduction, on-chip inductors often occupy a significant part of the chip area to meet performance requirements. This poses a major challenge in  RFICs, as increasing the inductance density and $Q$-factor while minimizing the chip area is necessary for advancing the technology. A possible solution to this problem is to design inductors leveraging the property of kinetic inductance in certain materials and configurations.

Kinetic inductance ($L_k$) is a property derived from the inertial mass of the electrons in the conducting material, which makes it oblivious to the area size of the inductor, unlike the standard Faraday magnetic inductance ($L_M$). The behavior of kinetic inductance can be examined using the standard Drude model for the frequency-dependent electrical conductivity, 
\begin{equation}
     \sigma(\omega) = \frac{\sigma_0}{1 + i \omega \tau},
 \end{equation}
 where $\tau$ is the relaxation time, which quantifies the damping of electron motion due to collisions. When the frequency-dependent term ($\omega \tau$) becomes non-negligible at high frequencies, the conductivity acquires a complex value. This reflects the finite time electrons require to accelerate under an applied force, and hence this inertial effect contributes to the impedance (inductance) of the material.In this regime, the impedance of a straight wire of length $l$ and cross-section $A$ can be expressed as \cite{kang2018chip},
\begin{equation} \label{eq:Lk}
 Z= \dfrac{l}{\sigma_0 A} \, (1+ i \omega \tau)= R_{DC}+ i \,\omega L_k,
\end{equation}
where the real part of $Z$ represents the DC resistance, and the additional imaginary part is identified as the kinetic inductive reactance, which depends on the cross-section of the wire and its total length, but is unrelated to the area of an inductive structure made of such wires. Equation (\ref{eq:Lk})  also applies in the regime $\omega \tau<1$, provided that the thickness of the wire is smaller than the skin-depth, ensuring that the current flows through the bulk of the wire.

From a circuit-level perspective, kinetic inductance behaves equivalently to magnetic inductance, with the kinetic energy of the electrons adding to the stored magnetic energy of the system \cite{khurgin2012reflecting}. This suggests that higher inductance densities and $Q$-factor values could potentially be achieved by exploiting kinetic inductance, without the need to increase the size of the inductor. The primary limitation of this idea, however, is that kinetic inductance is typically negligible in conventional metals at electronics-relevant frequencies, becoming significant only under specific conditions, such as in superconductors or at very high operating frequencies (typically, above the THz range). In both cases, the contribution of the motion of electrons to the total inductance can be substantial. While most research efforts on improving inductor performance have focused on optimizing the inductor layout to achieve larger $L_M$, a recent study \cite{kang2018chip} marked a potential paradigm shift by reporting a notable 1.5-fold increase in inductance density using intercalated graphene, a material engineered to exhibit relatively large kinetic inductance at radio frequencies. This study demonstrated the potential of carbon nanomaterials for next-generation inductors \cite{colmiais2022towards}. 

These recent advances offer an intriguing pathway for relaxing the upper bound on $Q$-factor discussed in the previous section. There are secondary effects that are not considered in the calculation of our numerical and analytical bounds, such as substrate losses, proximity resistance, and other coupling effects. If included, however, these effects can only result in a tighter bound, i.e., a lower $Q^m_{max}$ for specific configurations. Eq. (\ref{eq:Qm_max}) remains a valid upper bound when these effects are present. In contrast, the presence of kinetic inductance could potentially violate the derived upper bound, as it increases the total stored energy (of magnetic type) in the system. Thus, to generalize the bound, the analysis of $Q^m_{max}$ is  extended to account for the presence of kinetic inductance. In the MoM numerical analysis, this can easily be achieved by introducing a reactive component to the surface resistance $R_s$, modifying it to $Z_s = \dfrac{1}{\sigma_0 t}\,(1+i \omega \tau)$.

\begin{figure}[htbp]
\centering
\includegraphics[width=0.5\textwidth]{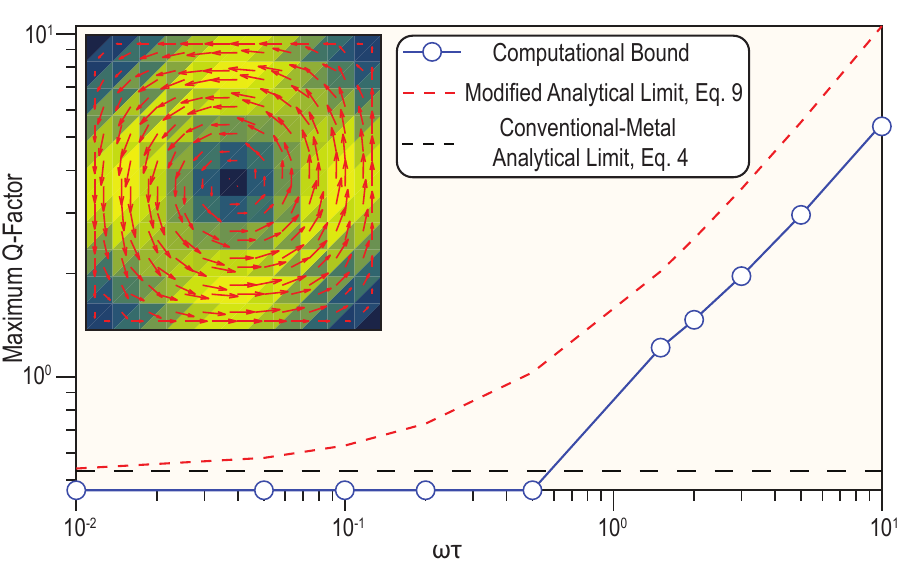}
\caption{\label{fig:fig3} Numerical results for the generalized bound, $Q^m_{max}$, which also accounts for the contribution of kinetic inductance, plotted for a square design area with $ka=1/1000$ considering a metal of conductivity $\sigma_0=2\times10^8$ S/m. The numerical bound (blue curve) exceeds the predictions of the analytical limit for conventional metals (\ref{eq:Qm_max}) (dashed black) in the regime where kinetic inductance becomes significant ($\omega \tau > 1$). In contrast, the modified analytical limit (\ref{eq:Qmax_Lk}) (dashed red) is a valid upper bound for the $Q$-factor in both the magnetic inductance-dominated region ($\omega \tau<1$) and the kinetic inductance-dominated region ($\omega \tau>1$). The inset shows the numerical solution for the optimal current distribution, after including the kinetic inductance contribution (the normalized plot looks the same for any $\omega\tau$ in the small $ka$ regime)}.
\end{figure}

Figure \ref{fig:fig3} shows $Q^m_{max}$ calculated using this modified surface impedance for a small optimizable square area with $ka=1/1000$, as a function of $\omega \tau$. Notably, the numerical solutions for $Q^m_{max}$ indeed significantly exceed the analytical bound (\ref{eq:Qm_max}) in the kinetic inductance regime ($\omega\tau>1$). The optimal current solution after including the kinetic inductance effect (inset of Fig. \ref{fig:fig3}) exhibits the same distribution as in Fig. \ref{fig:fig2}(c).I, namely, a single wide current loop. This behavior is expected since kinetic inductance is a material property and, therefore, its contribution to the $Q$-factor is independent of the inductor geometry. 

The resemblance of the optimal current distribution to that of a circular loop inductor facilitates a straightforward extension of the analytical bound (\ref{eq:Qm_max}) by adding the kinetic inductance contribution to the $Q$-factor, which leads to 
\begin{equation}\label{eq:Qmax_Lk}
 Q^m_{max}= \dfrac{\omega   (L_{loop}+L_k)}{ R_{loop}}\approx 0.15 \dfrac{ka}{R_s/\eta_0} + \omega \tau. 
\end{equation}
This modified bound (plotted in Fig. \ref{fig:fig3} with a red-dashed line) consists of two distinct terms. The first is a geometry-dependent term, which arises from the magnetic (Faraday) inductance and dominates for large values of $ka$. The second term is geometry-independent, arising from the kinetic inductance, and becomes dominant in the regime with large $\omega\tau$. Incidentally, we note that this result is consistent with the finding that optical resonances in small plasmonic structures, which are again governed by kinetic inductance, exhibit $Q$-factors that are determined solely by the relaxation time of the material and are, instead, independent of geometry \cite{khurgin2012reflecting,wang2006general}.

The extended bound given by Eq. (\ref{eq:Qmax_Lk}) can serve as a useful benchmark for the development of next-generation on-chip inductors that rely on the kinetic inductance effect. For example, Ref. \cite{kang2018chip} demonstrated a substantial improvement in the $Q$-factor of a square two-turn planar inductor of side length $200~\mu\text{m}$, increasing it from around $8$ to $12$ at $30$ GHz, by using intercalated-graphene instead of copper. Although this is indeed a significant achievement, our bound (\ref{eq:Qmax_Lk}) predicts that $Q$-factors up to $46$ could be attainable for the same area, suggesting that further design improvements may be possible.

\section{Conclusion}\label{sec13}
Determining fundamental performance limits can guide the development of advanced electromagnetic devices. While lower bounds on the $Q$-factor have been extensively investigated, especially in the context of efficient broadband antenna systems, the complementary problem of identifying the maximum possible $Q$-factor has been seemingly overlooked in the literature. In this paper, we have addressed this problem focusing on the technologically important goal of maximizing the $Q$-factor of planar inductors, which could benefit a wide range of applications, such as next-generation RFICs, RFID sensors, and miniaturized wireless power transfer systems.

Specifically, we have employed rigorous optimization methods to determine fundamental upper bounds on the $Q$-factor of electrically-small planar inductors. Approximate analytical expressions for this bound have also been presented. The derived fundamental limits are general, i.e., independent of any specific layout of the inductor, and depend only on conductor losses and the available design area. Optimal current distributions that attain the highest possible $Q$-factor, for all possible planar inductor designs, have been evaluated using convex optimization and modal analysis techniques based on the Method of Moments (MoM). By accounting for both radiation and ohmic losses, we have identified and physically interpreted nontrivial trends in the maximal $Q$-factor and optimal current distribution as functions of area and conductivity. Radiation loss could become a significant issue for inductor design at very high frequencies, such as millimeter-wave and sub-terahertz bands, making our results and insights particularly relevant in this regime. Our findings also show that maximal $Q$-factor density is associated with compact, symmetric geometries such as square or circular designs.

Although the optimal current distributions may be unrealistic to perfectly excite in practice, the resulting upper bound $Q^m_{max}$ and the associated physical insights can guide inductor design and provide strict reference limits against which practical designs may be assessed, as shown in Table \ref{tab:inductor_bounds}. The effect of kinetic inductance has also been considered, highlighting the potential of 2D materials with high relaxation times, such as intercalated graphene, to exceed the $Q$-factors achievable with conventional metals. This material-enabled effect can be particularly useful for small-scale designs where the geometrical degrees of freedom are limited, thereby contributing to the miniaturization of next-generation ICs.

For future studies, several additional factors could be incorporated, such as substrate losses, parasitic capacitances, and coupling to nearby inductors, to derive tighter and more practically relevant bounds for specific configurations of interest. The analysis could also be extended to three-dimensional inductor geometries. Moreover, future investigations could move beyond the single material assumption adopted here and explore heterogeneous designs. Notably, recent theoretical work has shown that, in some cases, multi-material systems can outperform physical bounds derived under a single-material assumption \cite{amaolo2024can}. Another important direction would be to investigate whether an optimal Pareto front exists that jointly maximizes both the inductance $L$ and the inductor $Q$-factor, a question of particular relevance to RF and power-electronics applications. 

More broadly, our work may open a path toward a rigorous treatment of $Q$-factor maximization across a wide range of RF, microwave, and millimeter-wave applications, while also inspiring analogous approaches for high-$Q$ optical systems \cite{puckett2021422,fang2024million}.

\appendices
\section{Method of Moments formulation}
Using the MoM formalism, the power dissipated within the system can be expressed in terms of the positive-definite resistance matrix $\bf{R}$ as, 
\begin{equation}
    P=\dfrac{1}{2} \bf{I}^H \bf{R} \bf{I},
\end{equation}
where $\bf{I}^H$ is the adjoint (conjugate transpose) of $\bf{I}$. The total energy stored in the system can be expressed as 
\begin{equation} \label{eq:Wtot}
    W= \dfrac{1}{4} \bf{I}^H \bf{X'}  \bf{I},
\end{equation}
where $\bf{X'=\partial \bf{X}/\partial\omega} $ is the derivative of the imaginary part of the impedance matrix of the system with respect to the operating frequency $\omega$ \cite{schab2018energy}. The total stored energy can also be expressed as the sum of the contributions from the stored magnetic and electric energies, i.e., $W=W_m + W_e$. In MoM terms, these energy components can be written as: 
\begin{equation}
    W_{\{m,e\}}= \dfrac{1}{4 \omega} \bf{I}^H \bf{X_{\{m,e\}}}  \bf{I},
\end{equation}
where $\bf{X_{\{m,e\}}} = \dfrac{\omega}{2}\, \big(X'\pm\dfrac{\bf{X}}{\omega}\big)$ are positive-definite matrices that represent the magnetic and electric energy contributions \cite{gustafsson2016antenna}.   In contrast, $\bf{X}$ can be negative-definite since it represents the overall reactance matrix of the system.  

For reliable circuit performance, particularly in matching networks, it is important to maintain the expected impedance characteristics of the inductors over the operational bandwidth, namely, the inductor structure should behave as a purely inductive element with impedance $Z(\omega)=j\omega L$. As can be deduced from the expression of the stored magnetic energy in current loops carrying constant current (e.g., Eq. (75) in Ref. \cite{vandenbosch2010reactive}), the inductance dispersion is negligible for small loops.  To minimize unwanted parasitic effects, the operation region of inductors is therefore usually chosen to be well below the inductor self-resonance frequency. Resonant frequency regions, characterized by comparable levels of stored energies ($W_m \approx W_e$), are also associated with significant ohmic losses. Consequently, this study focuses on the design of nondispersive inductors that are much smaller in size than the free-space wavelength at the operating frequency, which in turn is smaller than the self-resonance frequency of the structure. When this condition is met in inductive structures, the electric stored energy is smaller than the magnetic energy   ($W_m>W_e$). The frequency-dependent quality factor of the inductor can then be written as \cite{gustafsson2019tradeoff} 
\begin{equation} \label{eq:Qm}
Q^m= 2\omega \frac{W_m}{P}= \dfrac{ \bf{I}^H \bf{X_m} \bf{I}}{ \bf{I}^H \bf{R} \bf{I}},
\end{equation}
where $\omega$ is the operating frequency, not the self-resonance frequency, as usually done in the definition of $Q$-factor for reactive circuit components. 

The MoM formulation was originally developed to analyze perfect conductors,  where radiation is the only mechanism of energy dissipation. In practice, however, inductors are constructed from non-ideal metals with finite conductivity ($\sigma_0$), and non-zero skin depth ($\delta$). Furthermore, if the structure's dimensions are deeply subwavelength, ohmic losses dominate over radiation losses \cite{shahpari2018fundamental}. Since inductors are typically much smaller in size than the operating wavelength, accurately accounting for ohmic losses is crucial when calculating the upper bound on the $Q$-factor ($Q^m_{max}$) to ensure a tight bound. 
For practical, low-loss conductors, the MoM can be extended to account for ohmic losses using the concept of surface impedance $R_s$ \cite{kerr1999surface}. In this extension, the resistance matrix can be written as ${\mathbf{R}}={\bf{R_r}}+ R_s \bf{\Psi}$. Here, $\bf{\Psi}$ represents the Gram matrix associated with the MoM basis functions, which is typically diagonal or near-diagonal and can be straightforwardly evaluated \cite{gustafsson2019tradeoff}. The total power dissipated $P$ can then be divided into two components: the radiated power $P_r= \frac{1}{2} \bf{I}^H \bf{R_r}  \bf{I}$ and the ohmic power $P_l= \frac{1}{2} R_s\, \bf{I}^H \bf{\Psi}  \bf{I}$.

The surface impedance $R_s$ depends on the relative thickness of the conductor ($t$) with respect to the skin-depth ($\delta$). Under the assumption that ohmic losses are uniformly distributed along the conductor length, the surface impedance can be expressed as \cite{kang2018chip,gustafsson2019tradeoff}
\begin{equation}
  R_s = \dfrac{1}{\sigma_0\,  \min(\delta, t)} \,\,\,\     \text{(Ohm/square),}
\end{equation}
where $\sigma_0$ represents the metal DC conductivity and the skin-depth is  $\delta^{-1}=\sqrt{\omega\sigma_0\mu_0/2}$, with $\mu_0$ being the free-space permeability. In cases where the conductor thickness is much greater than the skin depth ($t \gg \delta$), an additional reactance component should be included in the expression of the surface impedance. Physically, this reactance component is due to the internal magnetic fields in the conductor's skin depth. However, this contribution is typically negligible, especially at high frequencies where the skin depth vanishes \cite{balanis2015antenna}. This internal inductance is different from the kinetic inductance effect discussed in the main text
, which stems from electrons' inertia rather than field penetration.

It is also important to note that the above formula for the surface impedance is inaccurate in the resonant regime of comparable thickness and skin-depth (see, e.g., Ref. \cite{kerr1999surface} for a detailed discussion). Moreover, the assumption of a uniform current distribution across the conductor width, which simplifies the modeling process in the standard 2D MoM formulation, can lead to a slightly larger estimate of $Q^m_{max}$ (hence, a slightly loose bound) due to neglecting the ohmic losses associated with the horizontal skin depth that could exist in wide conductors.

In this study, $\bf{X_m}$ is guaranteed to be positive-definite, since we are dealing with small, low-loss structures, which ensures that the magnetic stored energy is well defined. This is in contrast with the case of larger antennas, which are typically characterized by strong radiation and low $Q$-factors. In these cases, the previously mentioned definitions of stored energies may not be strictly valid due to the presence of significant losses and dispersion, which might even produce negative results. These challenges have been extensively studied in the literature on antenna optimization \cite{gustafsson2016antenna,schab2018energy}. 

Compared to $\bf{X_m}$, the matrix $\bf{R_r}$ is more difficult to work with as it is positive semi-definite, which means it has many zero and near-zero eigenvalues. Numerical inaccuracies could transform it into an indefinite matrix with negative eigenvalues, which could lead to nonphysical results. However, the positive semi-definiteness of $\bf{R_r}$ can be restored by replacing any negative eigenvalue with a zero value using matrix eigendecomposition \cite{gustafsson2016antenna}. Additionally, for the definitions of stored energy to remain valid, the frequency dependency of $\bf{R}$ must be negligible \cite{gustafsson2014q}, an assumption that is justified for the small low-loss structures considered in this study.  \\

\section{Determining the global maximum through convex optimization methods}
With all the necessary matrices defined, the maximization problem of (\ref{eq:Qm}) can now be established. An important distinction between the optimization problems for $Q_{max}$ and $Q_{min}$ should first be noted. In radiating systems that aim to achieve both a minimum radiation $Q$-factor. $Q_{min}$, and high radiation efficiency, the goal is to maximize radiated power while minimizing ohmic losses simultaneously. This dual goal results in a complex multi-objective optimization problem \cite{gustafsson2019tradeoff}. In contrast, for efficient inductive systems that require $Q^m_{max}$, the primary goal is to minimize all forms of power dissipation: both radiation and ohmic dissipation. This requirement can be incorporated into a single constraint in the optimization problem to find $Q^m_{max}$ as: 
\begin{equation} \label{eq:opt}
\begin{aligned}
    & \max_{\mathbf{I} \in \mathbb{C}^N} \quad \mathbf{I}^H \mathbf{A} \mathbf{I} \\
    & \text{subject to} \quad \mathbf{I}^H \mathbf{B} \mathbf{I} = c,
\end{aligned}
\end{equation}
where $\mathbf{A}=\mathbf{X_m-X_e}$, $\mathbf{B}=\mathbf{R}$, and $c$ is an arbitrary positive constant. The goal of (\ref{eq:opt})  is to find the optimal current solution $\mathbf{I}_{opt}$ that exhibits $Q^m_{max}$. The choice to maximize $\mathbf{I}^H (\mathbf{X_m-X_e}) \mathbf{I}$, rather than only $\mathbf{I}^H\mathbf{X_m}\mathbf{I}$, is to ensure that the excited structure is purely inductive. Maximizing $\mathbf{I}^H\mathbf{X_m}\mathbf{I}$ alone can result in resonant structures in which the stored electrical energy is also significant, which is not the focus of this paper.

As mentioned in the main text, Eq. (\ref{eq:opt}) represents a quadratically constrained quadratic program (QCQP) that is non-convex because the equality constraint is quadratic, which does not conform to the standard form of convex optimization \cite{boyd2004convex}. 
The procedure outlined next shows how the problem of finding $Q^m_{max}$ can be rigorously solved through the method of semidefinite relaxation (SDR). 

The first step is to introduce a Hermitian, positive-semidefinite, rank-1 matrix $\mathbf{\Pi=I\, I^H}$. This allows reformulating (\ref{eq:opt})  as linear functions of $\mathbf{\Pi}$ (and therefore convex), by leveraging  the properties of the trace operator ($\text{Tr}$) \cite{luo2010semidefinite}, leading to
\begin{equation} \label{ref:sdr}
\begin{aligned}
    & \max_{\mathbf{\Pi} \in \mathbb{H}^{N\times N}}  \quad \text{Tr}(\mathbf{A\, \Pi})\\
    & \text{subject to} \quad \text{Tr}(\mathbf{B\, \Pi}) = c \,\,\,\, \,\,\,\,\&\,  \,\,\,\,\,\,\mathbf{\Pi} \succeq 0\,\,\,\, \,\,\,\,\&\,  \,\,\,\,\,\,\  \mathrm{rank}(\mathbf{\Pi})=1
\end{aligned}
\end{equation}
where $\mathbf{\Pi}\succeq 0$ ensures that $\mathbf{\Pi}$ is positive semidefinite.  The only nonconvex element remaining in this formulation is the rank-1 constraint, as rank-1 matrices do not form a convex set. Finally, by relaxing this constraint, hence the name of the method, problem (\ref{ref:sdr}) becomes a standard semidefinite program.

Interestingly, it has been demonstrated that for problems with up to three constraints, such as the one considered here, solving the SDR problem is equivalent to solving the original non-convex problem, namely, it yields the same globally optimal solution\cite{luo2010semidefinite}. This fact also guarantees the existence of a rank-1 solution for $\mathbf{\Pi}$, from which the original optimal solution $\mathbf{I}_{opt}$ can be directly recovered \cite{luo2010semidefinite,jonsson2017methods}. It is important to note that sometimes numerical inaccuracies may result in solutions for $\mathbf{\Pi}$ with rank $> 1$. In this case, a rank-1 approximation can be applied to $\mathbf{\Pi}$ by selecting the eigenvector corresponding to the largest eigenvalue as an approximation of $\mathbf{I}_{opt}$.

\section{Characteristic Modes Theory} Although convex optimization guarantees the identification of a true bound on the $Q$-factor, as previously discussed, $Q^m_{max}$ can also be efficiently estimated, at least approximately, by applying the characteristic modes theory (CMT) to the components of the matrix $\mathbf{Z}$. CMT has been widely applied in antenna design problems, including for the problem of finding $Q_{min}$ \cite{cabedo2007theory,capek2016optimal,elias2021review}.  $Q^m_{max}$ can be estimated using CMT by solving the generalized eigenvalue problem that simultaneously diagonalizes $\mathbf{X_m-X_e}$ and $\mathbf{R}$. The highest positive eigenvalue $\lambda_1$ represents the modal solution that maximizes the ratio of net inductive stored energy  ($W_m>>W_e$) to the total power dissipated. Normalizing the eigencurrents, $\bf{I^H I}=1$, the eigenvalue  $\lambda_1 \approx\bf{I}^H \bf{X_m}  \bf{I}  / \bf{I}^H \bf{R}  \bf{I}= 2 \omega W_m / P $,  is exactly the $Q$-factor of a magnetic mode. 
Although in the general case the optimal current distribution $\mathbf{I}_{opt}$ for $Q^m_{max}$ is likely a combination of multiple modes of $\mathbf{Z}$, identifying the mode associated with $\lambda_1$ typically provides a highly accurate estimate of $Q^m_{max}$, particularly in electrically-small structures. As shown in Fig. \ref{fig:fig1}, virtually identical results are obtained using SDR and CMT methods.

The primary advantage of using the CMT method to estimate the optimal current $\mathbf{I}_{opt}$ lies in its computational efficiency. While the SDR method solves for $\bf{\Pi}$, of size $N \times N$, to find $\mathbf{I}_{opt}$,  the CMT method directly solves for $\mathbf{I}_{opt}$ (of size $N$). This reduction in complexity makes CMT a very practical approach, especially when a close approximation of the bound is sufficient. In the literature, CMT has yielded accurate results for $Q_{min}$ calculations \cite{capek2016optimal}. 

Finally, it is important to stress that both convex optimization and modal analysis methods presented in this manuscript rely on optimizing the current induced within the specified design area rather than optimizing the geometry of the conductor itself. While this approach allows formulating the optimization problem and determining true bounds in a relatively straightforward manner, it does not directly specify how to excite this optimal current or the exact inductor design that achieves $Q^m_{max}$. Exciting the optimal current distribution exactly is often an unrealistic goal due to practical limitations. Nevertheless, the presented approach guarantees that, for all possible designs within the given area, no other current distribution can outperform the computed optimal current with respect to the chosen metric. Furthermore, understanding the characteristics of the optimal current can guide the design of appropriate structures and feeding networks to excite near-optimal currents that achieve performance levels close to the established bound \cite{jelinek2016optimal,shi2017antenna,jonsson2018bounds,li2018increasing,capek2020finding}.

\section{Optimization algorithms} Both the SDR and CMT methods used in this work have been implemented using MATLAB \cite{MATLAB}. The core code implementing the MoM for computing the impedance matrix $\bf{Z}$ and its associated matrices is adopted from \cite{makarov2002antenna}, and is further developed to account for conductor losses. To solve the convex optimization problem of $Q^m_{max}$, we use the `\verb|CVX|' toolbox in MATLAB \cite{grant2014cvx}. For better convergence of the `\verb|CVX|' algorithm, it is recommended to use the square roots of the matrices, as outlined in Ref. \cite{gustafsson2016antenna}. The generalized eigenvalue problem is solved using MATLAB's built-in `\verb|eig|' function. 

\section*{Acknowledgment}
We would like to express our gratitude to Dr. Mohamed Ibrahim (Cornell
University) for valuable discussions. We also acknowledge support from the Air Force Office of Scientific Research with grant no. FA9550-22-1-0204 through Dr. Arje Nachman.

\ifCLASSOPTIONcaptionsoff
  \newpage
\fi

\bibliographystyle{IEEEtran}
\bibliography{bibtex/bib/Bibliography}


\begin{IEEEbiographynophoto}{Mohamed Ismail Abdelrahman}
(Graduate Student
Member, IEEE) was born in Mecca, Saudi Arabia.
He received the B.Sc. degree in electrical and
communication engineering from Ain-Shams University, Cairo, Egypt, in 2013, and the EUROPHOTONICS Erasmus Mundus Joint M.Sc. degree
from Aix-Marseille University, Marseille, France,
and Karlsruhe Institute of Technology, Karlsruhe,
Germany, in 2016. In 2026, he received a
Ph.D. degree in Electrical and Computer Engineering from Cornell University, Ithaca, NY, USA.
His current research interests include applied electromagnetism, metamaterials, quantum circuits, and silicon photonics.

\end{IEEEbiographynophoto}

\begin{IEEEbiographynophoto}{Matteo Ciabattoni}
(Graduate Student Member, IEEE) received the B.S. degree in Electrical Engineering from the University of Michigan, Ann Arbor in 2022 and the M.S. degree in Electrical and Computer Engineering from Cornell University in 2024, where he is currently pursuing the Ph.D. degree in electrical and computer engineering. In 2024 he was awarded the IEEE AP-S Doctoral Research Grant for his work in time-varying microwave circuits.
His research interests include microwave periodic structures, time-varying and nonlocal circuits.
\end{IEEEbiographynophoto}

\begin{IEEEbiographynophoto}{Francesco Monticone} (S’09–M’17–SM’26) is an Associate Professor in the School of Electrical and Computer Engineering at Cornell University, where he also currently serves as Director of Graduate Studies. He received the B.Sc. and M.Sc. (summa cum laude) degrees from Politecnico di Torino, Italy, in 2009 and 2011, respectively, and the Ph.D. degree in Electrical and Computer Engineering from The University of Texas at Austin in 2016, where he was advised by Prof. Andrea Alù. Dr. Monticone joined the faculty of Cornell University in 2017.
Dr. Monticone has authored over 150 peer-reviewed publications and delivered more than 80 invited talks and seminars. His research focuses on applied electromagnetics, metamaterials and metasurfaces, and nanophotonics, with a focus on theoretical foundations and novel wave phenomena. His work is motivated both by fundamental scientific questions about engineered wave-matter interactions and novel applications in a range of areas including imaging and sensing, information processing, communication, defense, and energy.
Dr. Monticone’s research and teaching have been recognized with several awards and honors, including the Cornell Engineering Research Excellence Award, the ONR Young Investigator Program Award (YIP) from the U. S. Department of the Navy, Office of Naval Research, the Franco Strazzabosco Award for Research in Engineering, the Michael Tien ’72 Sustained Excellence and Innovation in Engineering Education Award from Cornell University, the Leopold B. Felsen Award for Excellence in Electrodynamics, the AFOSR Young Investigator Program Award (YIP) from the U.S. Air Force Office of Scientific Research, and the Inaugural Margarida Jacome Dissertation Award from The University of Texas at Austin. Dr. Monticone is a member of the IEEE, the American Physical Society (APS), the Optical Society of America (Optica), The International Society for Optics and Photonics (SPIE) and has been elected a full member of the International Union of Radio Science (URSI).
Dr. Monticone regularly serves on technical program committees and steering committees, and as session chair and session organizer, for the leading conferences in photonics, electromagnetics, and metamaterials. He previously served as an Associate Editor of the IEEE Transactions on Antennas and Propagation and currently serves as an Associate Editor of Optica, the flagship journals of the IEEE AP-S Society and the Optica Society, respectively. He also served as the Technical Program Committee Chair for Metamaterials'2025.

\end{IEEEbiographynophoto}

\end{document}